\documentclass[useAMS,usenatbib]{mn2e}
\usepackage{ upgreek } 
\usepackage{graphicx}
\usepackage{amsmath}
\usepackage{float}
\usepackage{color}
\usepackage{rotating}

\graphicspath{ {figs/} }

\title[Eclipsing negative-parity image in microlensing]{Eclipsing negative-parity image of gravitational microlensing by a giant-lens star}
\author[Sohrab Rahvar ]{Sohrab Rahvar$^{ }$\thanks{E-mail: rahvar@sharif.edu}\\
$^{ }$Physics Department, Sharif University, P.O.Box 11365-9161, Azadi Avenue, Tehran, Iran}

\pagerange{\pageref{firstpage}--\pageref{lastpage}} \pubyear{}

\def\LaTeX{L\kern-.36em\raise.3ex\hbox{a}\kern-.15em
    T\kern-.1667em\lower.7ex\hbox{E}\kern-.125emX}

\begin{document}
\label{firstpage}

\maketitle

\begin{abstract}
Gravitational Microlensing has been used as a powerful tool for astrophysical studies and exoplanet detections. In the gravitational microlensing, we have two images with negative and positive parities. The negative-parity image is a fainter image and is produced at a closer angular separation with respect to the lens star. In the case of a red-giant lens star and large impact parameter of lensing, this image can be eclipsed by the lens star. The result would be dimming the flux receiving from the combination of the source and the lens stars and the light curve resembles to an eclipsing binary system. In this work, we introduce this phenomenon and propose an observational procedure for detecting this eclipse. The follow-up microlensing telescopes with lucky imaging camera or space-based telescopes can produce high resolution images from the events with reddish sources and confirm the possibility of blending due to 
the lens star. After conforming a red-giant lens star and source star, we can use the advance photometric methods and detect the 
relative flux change during the eclipse in the order of $10^{-4}-10^{-3}$.
 Observation of the eclipse provides the angular size of source star in the unit of Einstein angle and combination of this observation with the parallax observation enable us to 
calculate the mass of lens star.  Finally, we analyzed seven microlensing event and show the feasibility of observation of this effect in future observations. 
 \end{abstract}

\begin{keywords}
gravitational lensing:micro.
\end{keywords}

\section{Introduction}
Gravitational lensing referres to the bending of light by an astrophysical or cosmological object. This phenomenon has been 
predicted by theory of general relativity and later was confirmed in the eclipse of 1919 \citep{Einstein}. The gravitational lensing can be categorized into three classes of strong lensing, weak lensing and microlensing \citep{lensing}. The observations of each type of lensing provides information about the distribution of lens mass and this is a unique tool for investigation of dark matter distribution in the cosmological scales. The observation of gravitational lensing  in the Bullet cluster provides a strong evidence for the existence of dark matter  \citep{bcluster}. The gravitational lensing also has been studied on the map of Cosmic Microwave Background Radiation (CMB) to measure the foreground distribution of large scale structures \citep{planck}.

In the Milky Way scale where in the lensing system both lens and source are stars, the image separation from the lensing is in the order of micro-arc second where they can not be resolved from the ground-based  telescopes. This kind of lensing is so-called the gravitational microlensing.  The observational result from this lensing is the magnification of light of the source object \citep{pac}. This method has been used for detection of possible MACHOs \footnote{Massive Astrophysical Compact Halo Objects} in the Galactic halo and the observations of EROS, MACHO and OGLE groups for almost one decade in the direction of Large Magellanic Clouds revealed that MACHOs do not contribute in the dark matter budget of Galactic halo \citep{tiserand,ogle,moniez}. Also in recent years this method is used as a strong tool for detection of Earth-mass extrasolar planets beyond the snow line \citep{mao,gould,gaudi}. 
The observation of gravitational microlensing events has also important impact in astrophysical studies of distant 
stars \citep{rahvar1} and also can be used for amplifying signals from  Extraterrestrial intelligent life in SETI \footnote{Search for Extraterrestrial Intelligent Life} program \citep{rahvar2}. 

In the gravitational microlensing, the transverse velocity of lens, distance and mass of lens are degenerated parameters and the Einstein crossing time as the observable parameter is a combination of these three parameters. However, using the second order effects, we can partially break the degeneracy between the parameters of lensing. The two important effects are the parallax and finite-size of the source star. Here in this work we propose a new higher order effect that can provide extra information about the lensing system. 

In the gravitational lensing we have the two positive and negative parity images where the positive image is larger and is produced outside of the Einstein ring and the negative parity image is smaller and is produced close to the lens position. For the case of a red-giant lens star and large impact parameter, the negative parity image can be eclipsed by the lens star. We introduce this  phenomenon and discuss about the observability of it with the present telescopes. This observation can provide extra information about the lensing system. 

In section (\ref{introduction}) we introduce the basis of gravitational lensing and introduce the negative-parity image and possibility of eclipsing this image by a giant lens. In section (\ref{lightcurve}), we calculate the light curve of lensing during the negative-parity image eclipsing. In Section (\ref{estimate}) we study the feasibility of 
this effect by analyzing the light curve of seven microlensing light curve. Also we discuss about the future observations of this phenomenon. Finally we summarize this work in section (\ref{conc}).



\section{Basis of Gravitational microlensing} 
\label{introduction}

The formalism of gravitational microlensing can be done either in the geometric optics \citep{pac} or in the wave optics \citep{deg86} and even more fundamental formalisms as the Feynman path integral 
 \citep{pathi}. In the wavelengths shorter than the size of Schwarzschild radius, we can adapt the geometric optics and the lens equation in this limit reduces to  
\begin{equation}
\theta^2 = \beta\theta + 1, 
\label{lens}
\end{equation}
where $\theta$ is angular position of the image and $\beta$ is the angular position of the source. All the angles are measured with respect to the position of lens and they are normalized to the Einstein angle 
 \begin{equation}
 \theta_E = \sqrt{\frac{4GM_L}{c^2} \frac{D_{LS}}{D_SD_L}},
 \label{Ea}
 \end{equation}
where $D_S$, $D_L$ and $D_{LS}$ are the observer-source, observer-lens and lens-source distances, respectively and $M_L$ is the mass of lens. Solving equation (\ref{lens}), it has two solutions of 
\begin{equation}
\theta^\pm = \frac12(\beta \pm \sqrt{\beta^2 + 4}),
\end{equation}
where $\theta^\pm$ is the position of images form at the either sides of the lens position and are called the positive and negative parity images. The positive image form outside the Einstein ring and makes a larger image and the negative image reside inside the Einstein ring and forms a smaller image from the source. For the large impact parameters (i.e. $\beta\gg 1$) the negative image approaches to the lens position. The separation between the lens and images for negative and positive parity images are
\begin{eqnarray}
\label{nmap}
|\theta^{-}| &=& \frac{1}{\beta} - \frac{1}{\beta^3}+\frac{1}{\beta^5} ... \\
\theta^+ &=& 1 + |\theta^{-}|,
\label{c1}
\end{eqnarray}
where for a large impact parameter, ignoring the higher order terms results in $|\theta^{-}|  = {1}/{\beta}$.

Now, we calculate the magnification of light due to lensing of source star, assuming that the source is a point like object. The magnification results from the Jacobian of transformation between the source and image, assuming that the surface brightness of images are the same as surface brightness of the source, 
\begin{equation}
A^\pm = |\frac{\theta^\pm}{\beta}\frac{\partial\theta^\pm}{\partial\beta}|,
\end{equation}
where for a large impact parameter which is in our concern, the magnification for positive and negative parity images are
\begin{eqnarray}
A^{+} &=& 1 + \frac{1}{\beta^4}, \\ 
A^{-} &=& \frac{1}{\beta^4},
\end{eqnarray}
and the overall magnification is $A=|A^+| + |A^-| = 1 + 2/\beta^4$.

The other important issue is the shape of images from the lensing. We assume a circular shape for the source star with  the angular size $\rho_\star$, normalized to the Einstein angle (i.e. $\rho_\star = R_S/(D_S\theta_E)$). For simplicity, we take the lens located at the centre of coordinate and centre of source located along the $x$- axis and distance of $\beta_0$ from the lens. All the angular scales are normalized to the Einstein angle. The equation of circle in the polar coordinate can be written as 
\begin{equation}
\beta^2 + \beta_0^2 - 2\beta_0\beta\cos\phi = \rho_\star^2.
\end{equation}
The map of this circle for the negative-parity image for a large impact parameter, using equation (\ref{nmap}) in the cartesian coordinate is also a circle 
\begin{equation}
 (\theta_x^{-}+ \frac{\beta_0}{\beta_0^2 - \rho_\star^2})^2 + {\theta_y^{-}}^2 = \frac{\rho_\star^2}{(\beta_0^2 - \rho_\star^2)^2},
\end{equation} 
where the radius of image is $\alpha^{-}_I = \rho_\star/(\beta_0^2 - \rho_\star^2)$ and centre of image is located at $\theta_0^{-} = -\beta_0/(\beta_0^2-\rho_\star^2)$. We note that for large impact parameter, $\rho_\star\ll \beta_0 $ then $|\theta_0^{-}| \ll \beta_0$ or in another word, the negative parity image is very close to the lens star.  Taking into account the condition of $\beta_0\gg \rho_\star $ the radius of image simplifies to $\alpha^{-}_I \simeq \rho_\star/\beta_0^2$ and $\theta_0^{-} \simeq - 1/\beta_0$. Let us assume $R_L$ as the radius of lens star, then $\alpha_L = R_L/(D_L\theta_E)$ is the angular radius of lens normalized to the Einstein angle. 

\begin{figure}
\centering
  \includegraphics[width=90mm]{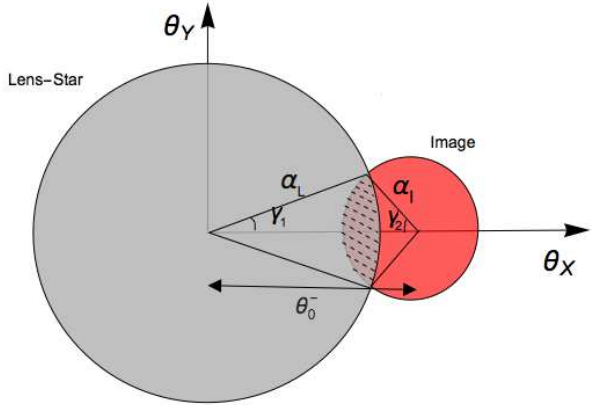}
  \caption{Schematic configuration of lens star (grey circle) with negative-parity image (red circle) beside the lens star. Here the image 
  is partially eclipsed by the lens star. The distance of centre of lens to the centre of negative-parity image is given by $\theta_0^{-}$. The radius of lens and image are $\alpha_L$ and $\alpha_I$, respectively. The angle between the centre of lens to the overlap region and line connecting the centres are
  labeled by $\gamma_1$ angle. We have the similar definition for $\gamma_2$ angle. All the angles are normalized to the Einstein angle and they are dimensionless. The hatched area of image is covered by the lens star.}
  \label{fig1}
\end{figure}

\begin{figure}
\centering
  \includegraphics[width=90mm]{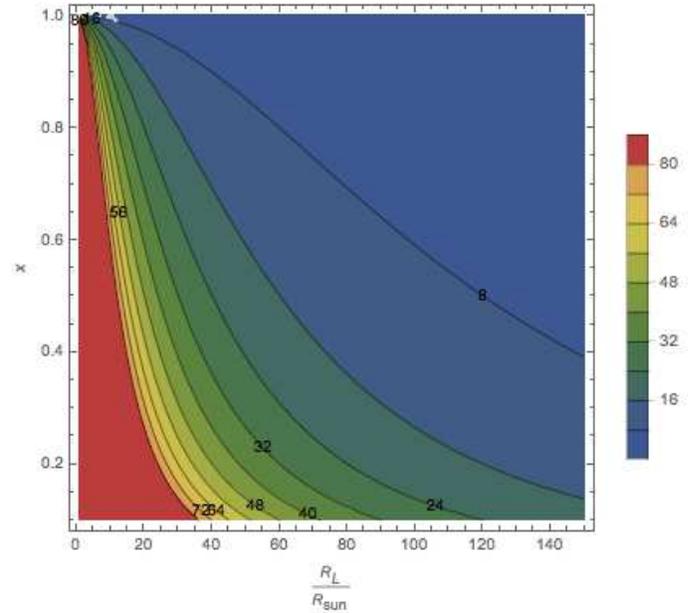}
  \caption{This figure illustrates the necessary condition of eclipsing (i.e. $\beta_0 = \alpha_L$). The numerical values of the contour represents $\beta_0$ as function of radius of lens star in the unit of solar radius and ratio of observer-lens distance to the observer-source distance (i.e. $x = D_L/D_S$). The source stars are located at the Galactic Bulge. }
  \label{fig2}
\end{figure}

Figure (\ref{fig1}) demonstrate the position of the negative-parity image compare to the lens star. For the large impact parameter, the eclipse for the image happens when $|\theta_0^{-}|\le {\alpha_L} + \alpha^{-}_I$ which implies $1/\beta_0\le \alpha_L + \rho_\star/\beta_0^2$. Then the geometrical condition for the ingress of eclipse is
\begin{equation}
\beta_0 \ge \frac{1}{2\alpha_L}( 1+ \sqrt{1-4\rho_\star\alpha_L}),
\label{ineq}
\end{equation}
We note that all the angles in this formula are normalized to the Einstein angle and they are dimensionless. 
For simplicity we use the definitions of $\alpha_L$, $\rho_\star$ and Einstein angle from equation (\ref{Ea}), and express these two parameters in simpler form of 
\begin{eqnarray}
\label{alpha1}
\alpha_L &=& \alpha_\odot \left(\frac{x}{1-x}\right)^{1/2}\left(\frac{M_L}{M_\odot}\right)^{-1/2}\left(\frac{R_L}{R_\odot}\right) \\
\rho_\star &=& \alpha_\odot x \left(\frac{x}{1-x}\right)^{1/2} \left(\frac{M_L}{M_\odot}\right)^{-1/2}\left(\frac{R_S}{R_\odot}\right), 
\end{eqnarray}
where $x=D_L/D_S$ and $\alpha_\odot = R_\odot/(D_L\theta_{E,\odot})$ is the angular size of solar-type lens, located at $x=1/2$ and normalized the Einstein angle with a solar mass lens. For the case of lenses towards the Galactic Bulge, the distance is $D_S = 8.5$kpc which results in $\alpha_\odot = 0.96 \times 10^{-3}$.  
Using the numerical values for radius of giant stars, expression (\ref{ineq}) can be simplified as 
\begin{equation}
\beta_0 \ge  {\alpha_L}^{-1}.
\label{cond}
\end{equation}
In order to have a numerical value for $\beta_0$, we take the radius of lens-star ranges from brown dwarf $R \simeq 7\times 10^4$ km \citep{bd} to AGB stars with $R \simeq 215 R_\odot$ \citep{agb} and hypergiants in the order of $R\simeq 1000 R_\odot$ \citep{hg}. Let us choose the typical mass of $M_L = 0.3 M_\odot$ for the lens and assuming microlensing observation is performed toward the Galactic Bulge. Figure (\ref{fig2}) represents $1/\alpha_L$ as a function of lens radius and $x = D_L/D_S$.  For a red-giant lens star located at the distance of $x=0.9-1$ which is a typical position of 
red clump giants, the impact parameter to have negative-parity eclipse is around $\beta_0 \simeq  5$.  This means that after almost 
five times of Einstein crossing time from the peak of light curve, we would observe the observational feature of eclipsing image
which results in decreasing the overall flux from the images and lens star. 
 
In what follows, we estimate the decrease of flux due to eclipse of negative image by a red giant lens star. More precise calculation is done in 
the next section. The total flux of source star and lens during the magnification can be written as 
\begin{equation}
F = (|A^+| + |A^-|)F_S + F_L,
\label{tot}
\end{equation}
where $F_S$ and $F_L$ are the flux of source and lens star and $F$ is the total flux during the magnification. During the eclipse the magnification term of source with negative parity image cancels and contrast in the overall flux of lens and source is given by 
\begin{equation}
\frac{\Delta F}{F} = - |A^-| \frac{F_S}{F }.
\label{def}
\end{equation}
We note that due to large impact parameter of lensing compare to the size of source star (i.e. $\beta_0\gg\rho_\star$), we use the point-like 
magnification relation in equation (\ref{def}). Substituting the total magnification from equation (\ref{tot}) and $A^-$ for large impact parameter, results in 
\begin{equation}
\frac{\Delta F}{F} = - 3.2\times 10^{-3} \left(\frac{b}{2}\right) \left(\frac{\beta}{5}\right)^{-4},
\label{df}
\end{equation}
 where $ b = F_S/(F_S + F_L) $ is the blending factor which results from the mixing the lens flux with the flux of source star. For the case 
 that both the source star and the lens star are red giants, $b\simeq 2$ and ${\Delta F}/{F}$ is in the order of $10^{-3}$. However, for a typical solar-type star as a 
 source and a red giant as a lens, the flux-contrast due to image eclipsing on the overall flux of source and lens would be in the order of $10^{-5}$. 
 
 We note that blending also can be due to the background stars that mix with the light of source star within the PSF \footnote{Point Spread Function} of source star. In recent years with development of imaging techniques as the lucky-imaging camera, we can produce high resolution images from the 
 microlensing events and compute correctly the blending parameter and the source of blending \citep{sajadian2016} or we may use space-based 
 telescope like Spitzer that recently is used for the microlensing observations \citep{spitzer}. Also for 
 high precision photometry,  in recent years, with developing telescope defocusing method,  the photometric reaches to milli-magnitude precision 
 for nearby stars in the field-stars \citep{south} and this method might be used in uncrowded fields of microlensing stars.

 
 \subsection{Time scales of Eclipsing}
 
The other observable parameter associated to the eclipsing of negative-parity image is the time scales of this phenomenon. We are interested in to know how long after the peak of microlensing light curve, we would expect to detect the ingress of eclipse of negative-parity image. We start with the impact parameter of lens which is given by $\beta_0^2 = \beta_{min}^2 + (t-t_0)^2/{t_E}^2$ and use equation (\ref{ineq}) which provides the geometrical condition of ingress and obtain the relative time of eclipsing compare to the peak of 
magnification time as follows
\begin{equation}
t_{ec} - t_0 = t_E\left[\frac{1}{4\alpha_L^2}(2 - 4\rho_\star\alpha_L +2\sqrt{1-4\rho_\star\alpha_L})-\beta_{min}^2 \right]^{1/2}.
\label{diff}
\end{equation}
Here for calculating the eclipsing time we need to know $t_E$ and $\beta_{min}$ from the best fit to the observed light curve, the
distance of lens from the observer and the lens mass which are measurable by combining the results of parallax and finite source effects, the radius of lens star and source star from the spectral analysis of event. 
 
The typical time scale for the Einstein crossing time of microlensing events toward the centre of Galaxy is in the order of one month \citep{ogleIII}. On the other hand from Figure (\ref{fig2}), for the lenses located at $x>0.9$, $1/\alpha_l$ is less than $10$. Then we expect that 
$t - t_0$ would be less than one year. Hence to observe this phenomenon, we need to perform a follow-up observation after almost year (for a typical event) from the peak of microlensing light curve to detect the eclipsing of negative-parity image.

The second time scale associated to the image eclipsing is the duration of eclipsing, $\Delta t = t_2 - t_1$ that takes from the start of image ingress at $t_1$ to the completion of eclipse of image at $t_2$. From equation (\ref{nmap}), the position of centre of image at time $t_1$ and $t_2$ are  $\theta_1^{-} = {1}/{\beta_1}$ and  $\theta_2^{-} = {1}/{\beta_2}$, respectively. Ignoring the minimum impact parameter compare to the impact parameter at the time of eclipsing (i.e. $\beta_{min}<\beta_0$), the impact parameter is related 
to time as $\beta = (t-t_0)/t_E$, where substituting in the duration of eclipsing  in which the centre of image displaces by the amount of diameter of 
image (i.e. $ \theta_1^{-} - \theta_2^{-} = 2\alpha_I^{-}$) and combining with equation (\ref{diff}) results in 
\begin{equation}
\Delta t = 2 \rho_\star t_E.
\label{duration}
\end{equation}  
Using the typical values for $t_E \simeq 30$ days and $\rho_\star\simeq 10^{-2}$, we obtain $\Delta t \simeq 0.6$ day. 
This means that for duration of one day, the overall flux of lens and source star will decline by the amount that is given 
by equation (\ref{df}) and for the case of being both lens star and source star as red giant, $\Delta F/F$ would be in the order 
of $10^{-3}$.
 
\section{Light Curve during eclipse}
\label{lightcurve}
In this section we calculate the light curve of flux receiving from both the lens and source stars during the eclipse of negative-parity image. We use the the geometrical configuration of Figure (\ref{fig1}) to calculate the area of image that is eclipsed by the lens star. The hatched area of 
Figure (\ref{fig1}) is given by $\Delta S = \alpha_L^2\gamma_1 + \alpha_I^2\gamma_2 - \theta_0^{-}\alpha_L\sin\gamma_1$ where $\gamma_1$ and $\gamma_2$ are the angles between the line connecting the centre of lens to the centre of image and lines from both centres connected to the 
edge of hatched area. We replace $\gamma_1$ and $\gamma_2$ in terms of the radius of two circles and distance between the centre of circles (i.e. $\theta_0^{-}$) and rewrite the hatched area in the following form 
\begin{eqnarray}
\Delta S &=& \alpha_L^2\cos^{-1}(\frac{\alpha_L^2 + {\theta_0^{-}}^2 - {\alpha_I^{-}}^{2}}{2\alpha_L\theta_0^{-}} ) + {\alpha_I^{-}}^2\cos^{-1}(\frac{{\alpha_I^{-}}^2 + {\theta_0^{-}}^2 - \alpha_L^{2}}{2\alpha_I^{-}\theta_0^{-}} ) \nonumber \\
&-&\frac12\left[(\alpha_L + \theta_0^{-})^2 - \alpha_L^2\right]^{1/2}\left[\alpha_L^2 -(\alpha_L-\theta_0^{-})^2 \right]^{1/2},
\label{hash}
\end{eqnarray}
where replacing with $\theta_0^{-} = \alpha_L + \alpha_I^{-} - y$  this equation simplifies to 
\begin{eqnarray}
\label{ds}
\Delta S &=& \alpha_L^2\cos^{-1}\left[\frac{2(\alpha_L+\alpha_I^{-})(\alpha_L-y) + y^2}{2\alpha_L (\alpha_L + \alpha_I^{-} -y)}\right]  \\
 &+& {\alpha_I^{-}}^2\cos^{-1}\left[\frac{2(\alpha_L+\alpha_I^{-})(\alpha_I^{-}-y) + y^2}{2\alpha_I^{-} (\alpha_L + \alpha_I^{-} -y)}\right] \nonumber \\
&-&\frac12\left[(2\alpha_L-y)(2\alpha_L + 2\alpha_I^{-} -y) \right]^{1/2}\left[2\alpha_I^{-} y-y^2\right]^{1/2}, \nonumber
\end{eqnarray}
in which $y = 2 \alpha_I^{-} \times (t-t_1)/{\Delta t}$ represents the displacement of image towards the lens, starting from 
the ingress point and ranges by $y\in[0,2\alpha_I^{-}]$.

 \begin{figure}
 \centering
 \includegraphics[width=90mm]{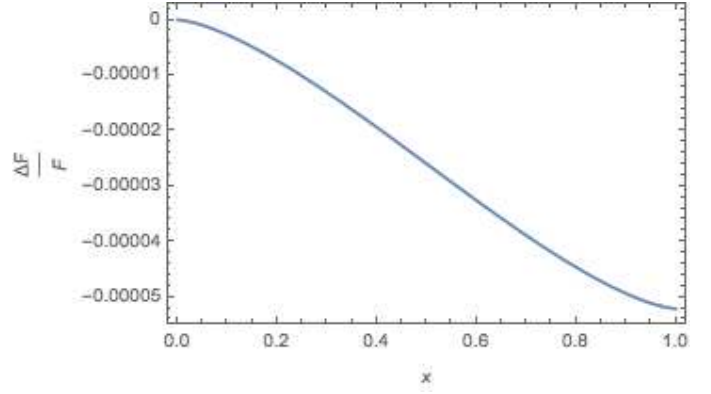}
 \caption{$\Delta F/F$ in terms of $x = y/2\alpha_I$ during the negative-parity eclipsing. The size of lens and source are chosen as
 $\alpha_L = 0.01$ and $\rho_\star = 0.01$ which corresponds to lens and source red giants with $20 R_\odot$ radius and Einstein 
 radius of lens with one astronomical unit.}
 \label{fig3}
 \end{figure}

In order to calculate the decline in the overall flux of source and lens, we simplify the problem by assuming that the spectral type 
of these two stars are almost similar and they have almost the same temperature. Then the flux of lens star and source star is 
only a function of size of these two objects and the contrast in the flux due to eclipse would be 
\begin{equation}
\frac{\Delta F}{F} = - \frac{\Delta S}{\pi(\alpha_L^2 + \rho_\star^2)},
\end{equation}
where $\Delta S$ is given by equation (\ref{ds}). We note that the size of image is expressed by the size of source star and impact parameter at the time of eclipse. We use equation (\ref{ineq}) and express the radius of image in terms of source radius and lens radius as follows
\begin{equation}
\alpha_I^{-} = \frac{4\rho_\star \alpha_L^2}{(1 + \sqrt{1-4\rho_\star \alpha_L})^2}.
\end{equation}
By substituting this equation in (\ref{ds}), we can conclude that the flux contrast during the negative-parity image eclipse depends on the 
size of source star, size of lens and proximity of the lens star to the image (i.e.   $\Delta F/F(\rho_\star,\alpha_L,y)$). Figure (\ref{fig3}) 
represents the flux contrast (i.e. $\Delta F/F$) during negative parity image eclipsing for the case of having red giant lens star and red giant source
star with the radii of $R = 20 R_\odot$ and Einstein radius in the order of astronomical unit (i.e.  $\alpha_L \simeq 0.01$ and $\rho_\star \simeq 0.01$).

From the observation of this light curve, we can derive the parameters of $\alpha_L$ and $\rho_\star$. One the physical parameters that we can extract from these parameters is the size of Einstein angle that needs having information about the size of source star. The size of source star from the position of star in the absolute colour-magnitude diagram can be estimated. The practical procedure to know the colour and magnitude of source star is the multicolour photometry. Let us assume the photometry of a microlensed event at least with two colours at the peak of event and at least at two different times. We assume using $V$ and $I$ band filters in this observation. The total flux of lens and source star ($F_{i}^{(T)}(t)$) as a 
function of time is composed of flux of source star ($F_{i}^{(S)}(t)$) and flux of lens star $F_i^{(L)}(t)$ as follows
\begin{equation}
\label{me}
F_{i}^{(T)}(t_j) = A(t)F_{i}^{(S)}(t_j) + F_i^{(L)}(t_j),
\end{equation}
where subscript $i$ represents different filters (in our case $i=V$ and $I$) and $j$ represents the observation of event at different moments, let us take $t_1$ and $t_2$. Then the total flux of lens and source star with solving four equations in (\ref{me}) results in 
\begin{eqnarray}
F_{i}^{(S)} &=& \frac{F_i(t_2) - F_i(t_1)}{A(t_2) - A(t_1)}, \\
F_{i}^{(L)}  &=& \frac{F_i(t_1)A(t_2) - F_i(t_2)A(t_1)}{A(t_2) - A(t_1)}. \nonumber 
\end{eqnarray} 
Moreover, the colour and magnitude of the source star also can be obtained in the high magnification events \citep{hmag} if the  
observation is performed in two different filters at the peak of light curve. In this case we can ignore the contribution of the blending effect in the colour and 
magnitude of the source star.  The spectroscopic observation of microlensing event can also provide more information about the precise 
position of the source star in the colour-magnitude diagram.

Using the Interstellar Extinction Calculator based on the OGLE webpage \footnote{2 http://ogle.astrouw.edu.pl/} we can  find the extinction for the source star \citep{nataf}. The result is the extinction correction of the colour and the apparent magnitude of the source star. Since there are relation between the surface brightness and color of stars, we can use the following relation that provides directly the angular size of 
source star in terms of the colour and the magnitude of the star \citep{pascal}
\begin{equation}
\log_{10}(\theta_\star) = 3.1982 + 0.4895(V-I)_0 - 0.0657(V-I)_0^2 -0.2 I_0.
\label{anglsize}
\end{equation}
For the bugle events, it is more likely that source star belong to the bulge of Galaxy which provides observer-source distance and that 
results in calculation of the radius of source star (i.e. $R_s$). On the other hand we have seen that from the light curve of eclipsing the parameter of $\rho_\star$ can be derived. The result would be the Einstein angle from knowing $D_S$, $R_S$ and $\rho_\star$ (i.e. $\theta_E = R_S/D_S\rho_\star$).



There is another second order effect in the microlensing light curve so-called parallax effect that provides extra information about the lens and 
source star \citep{gould1992,rahvar2003}. The parallax effect during the microlensing results from the orbital motion of Earth around the Sun and causes a relative non-zero acceleration between the angular motion of lens and source with respect to the Earth. The result is a small deviation in the light curve compare to the 
symmetric light curve of single lensing. The parallax parameter 
\begin{equation}
\pi_E = \frac{1 au.}{\theta_E}(\frac{1}{D_L} -\frac{1}{D_S}),
\label{parallax}
\end{equation}
is the relevant parameter in this effect and can be measured from the best fit to the observational data. One the other hand by measuring the Einstein angle from the negative-parity eclipsing, we can calculate the mass of lens by 
\begin{equation}
M_L = \frac{\theta_E}{\kappa\pi_E}
\label{mass}
\end{equation}
where $\kappa = {4G}/{1AU c^2} = 8.144 ~mas/M_{\odot}$. Measuring the mass of lens from the microlensing observations is one the important goals of this experiment to identify the mass function of the stellar and non-stellar objects as well as their distribution in the Galaxy.

\section{Feasibility of Observation of eclipsing negative-parity image} 
\label{estimate}

In this section we study the feasibility of observation of eclipsing negative-parity image of microlensing events. Since the time scale of eclipsing 
after the peak of event is almost one year and a precise photometry is needed for it, we have to calculate the time and duration of eclipsing in advance. 
In order to study the feasibility of prediction of eclipsing time scales,  we select a small subset of microlensing events from a large number of events towards the Galactic Bulge where both the parallax and finite--source parameters of these events 
have been measured. These parameters are essential in 
calculation of the two eclipsing time scales  of  $\Delta t$ and $t_{ec} - t_0$.  In our list of microlensing events, for 
some of events the parallax parameter is not well determined. However, using the Galactic 
model for the disk, from the likelihood function we can estimate the mass and the distance of lens from the observer.

Most of our events in this set are highly magnified events where the probability of the  
planetary signals in these events are high. These events are precisely observed by the survey and the follow-up telescopes, by means of 
the high cadence light curve and locating star in the colour-magnitude diagram. Table (\ref{table2}) shows the 
list of events we choose for analyzing and the associated parameters that have been 
measured from the light curve. In the last two column we determine the duration of eclipsing time and the starting time of eclipsing 
relative to the peaking time of the light curve. One of the essential parameters that has to be determined from the parameters of light curve
 is $\alpha_L$ which depends on the parameters of lens as follows
\begin{equation}
\alpha_L = 4.64\times 10^{-3} (\frac{R_L}{R_\odot})(\frac{\theta_E}{1mas})^{-1}(\frac{D_L}{kpc})^{-1},
\end{equation}
where $\theta_E$ is measured from finite source effect by estimating the radius of source star from the colour-magnitude 
diagram and distance of source star (i.e. $\theta_E = R_S/(\rho_\star D_S)$).  On the other hand from the parallax observation, we 
can obtain the mass of lens from equation (\ref{mass}) and subsequently the distance of the lens star. We estimate the radius of lens star 
either in the case of being a main sequence star \citep{book} or a dwarf \citep{mr}. Since we only measure the 
mass of lens star, the evolution of star unless a careful spectroscopic observation is done is not known. For large 
masses for the lens, we also identify its radius assuming that the lens is in the phase of a red-giant star \citep{rg},  
\begin{equation}
R \simeq \frac{3.7\times 10^3}{1 + \mu^3 + 1.75\mu^4}R_\odot
\end{equation}
where $\mu = M_L/M_\odot$ and the domaine of this parameter is $0.17<\mu<1.4$. In the calculation of eclipsing time scales, for those events 
that have larger lens mass, we assume the radius of lens star being in two cases of main-sequence or red-giant star.

\begin{table*}
\tiny
\begin{center}
\begin{tabular}{|c|c|c|c|c|c|c|c|c|c|}
\hline\hline
Name & $t_E$      & $\beta_{min}$ & ${\rho_\star} ({10^{-3}})$               &  $\theta_E$ & $D_L$   & $M_L$              & $\alpha_L (10^{-3})$        &    $\Delta t$  & $t_{ec} - t_0$ \\ 
 (1)     & (2)~(day) & (3)                  & (4)  &(5) ($mas$)  & (6) (kpc) & (7) ($M_\odot$)& (8)~&~(day)(9)     & ~(day)(10) \\      \hline
OGLE-2011-BLG-265 & $53.63^{+0.19}_{-0.19}$ & 0.13 & $9.73^{+1.89}_{-1.89}$ & $0.42^{+0.04}_{-0.04}$ & $4.4^{+0.5}_{-0.5}$ & $0.14^{+0.06}_{-0.06}$ & $0.47^{+0.09}_{-0.09}$ &  $1.04^{+0.20}_{-0.20}$ & $\sim113270$ \\ \hline
KMT-2015-1b~~~~~~~~~~ & $10.9^{+0.10}_{-0.10}$ & $0.221^{+0.003}_{-0.003}$ & $44.7^{+0.8}_{-0.8}$ & $0.098^{+0.013}_{-0.013}$& $8.2^{+0.9}_{-0.9}$ & $0.18^{+0.12}_{-0.12}$ &$1.15^{+0.19}_{-0.19}$&$0.97^{+0.09}_{-0.09}$ & $9438^{+1604}_{-1604}$\\  \hline
OGLE-2004-BLG-254 & $13.23^{+0.04}_{-0.05}$ & $4.6^{+0.76}_{-0.86}10^{-3}$ & $40.0^{+0.00}_{-0.02}$ & $\sim 0.114$ & $\sim 9.6$ & $\sim 0.11$ & $\sim 0.50$ &$\sim 1.05$ & $\sim 26000$\\  \hline
OGLE-2007-BLG-050 & $66.9^{+0.6}_{-0.6}$ & $0.002$ & 4.5 & $0.48^{+0.01}_{-0.01}$  & $5.5^{+0.4}_{-0.4}$ & $0.50^{+0.14}_{-0.14}$ & $0.59^{+0.35}_{-0.35}$  & $0.602^{+0.005}_{-0.005}$ & $\sim111952$\\  \hline
OGLE-2007-BLG-050(R) & --- & --- & --- & ---  & --- & --- & $\sim 328$  & --- & $\sim 187 $\\  \hline
OGLE-2008-BLG-290 &  $16.361^{+0.077}_{-0.077}$ & $0.00276^{+0.00020}_{-0.00020}$ & 22.  & $\sim 0.3$ & $\sim 6$ kpc & $\sim 0.3$ & 1.28 &  $\sim 0.7$& $\sim$ 12646\\  \hline
OGLE-2008-BLG-290 (R) &  --- & --- & 22.  & --- & ---& --- & $\sim 28$ &  --- & $\sim 225$ \\  \hline
OGLE-2008-BLG-279 &  $106^{+0.9}_{-0.9}$ & $6.6^{+0.5}_{-0.5}\times 10^{-4}$ & $0.68^{+0.06}_{-0.06}$  & $0.81^{+0.07}_{-0.07}$ & $4.0^{+0.6}_{-0.6}$& $0.64^{+0.1}_{-0.1}$ & $ 1.05^{+0.18}_{-0.18} $ &  $0.140^{+0.012}_{-0.012}$ & $100023^{+17303}_{-17303}$ \\  \hline
OGLE-2008-BLG-279 (R) & ---&---& --- & ---& ---&--- & $ 571^{+98}_{-98} $ &  ---& $185^{+32}_{-32}$ \\  \hline
MOA-2009-BLG-174 & $64.99^{+0.61}_{-0.61}$ & $0.0005^{+0.0001}_{-0.0001}$& $2.0^{+0.1}_{-0.1}$ & $0.43^{+0.04}_{-0.04}$& $6.39^{+1.11}_{-1.11}$& $0.84^{+0.37}_{-0.37}$ & $ 1.51^{+0.26}_{-0.26} $ &  $ 0.260^{+0.012}_{-0.012}$ &  $42761^{+7269}_{-7269}$\\ \hline 
MOA-2009-BLG-174(R) & --- & ---& ---& --- & ---& --- & $1262^{+214}_{-214}$ & --  & $51.3^{+8.7}_{-8.7}$ \\  

\hline \hline
\end{tabular}
\caption{Calculation of the time scales of eclipsing negative-parity microlensing event for seven event. The first column represents the name of event. The extension of $(R)$ in some of events is assuming the lens star is taken a red-giant star. The second column is the Einstein crossing time of event. The third column is the minimum impact parameter of event. The fourth column is the finite source parameter (i.e. $\rho_\star$), scaled by $10^{-3}$.  The fifth column is the Einstein angle in mili-arc second. The sixth column is the distance of lens in $kpc$ which is obtained either by parallax effect or likelihood function assuming a model for the distribution of matter in the Galaxy. The seventh column is the mass of lens scaled to the solar mass. The eighth column is $\alpha_L$ (angular size of lens star normalized to the Einstein angle). The ninth column is the duration of eclipse and the tenth column is the 
eclipsing time relative to the peaking time of the light curve. 
}\label{table2}
\end{center}
\end{table*}

We adapt the parameters of light curve of the following events of OGLE-2011-BLG-265 \citep{265}, KMT-2015-1b \citep{kmneta}, OGLE-2004-BLG-254
\citep{254}, OGLE-2007-BLG-050 \citep{50}, OGLE-2008-BLG-290 \citep{290}, OGLE-2008-BLG-279 \citep{279} and 
MOA-2009-BLG-174 \citep{hmag}. From analyzing these events, we obtain the time scale for the eclipsing which is in the order of one day and 
the eclipsing time relative to the peak of light curve which is very large for the case that the lens star is a main sequence or dwarf star and is in 
the order of hundred days when the lens star is a red giant. Identifying the spectral type of lens star is very important to explore 
a red-giant lens star and determine precisely the eclipsing time of the event.

\section{Conclusion}
\label{conc}
In this work, we proposed the observation of eclipse of negative-parity image from single-lens microlensing events. The gravitational lensing produces two images at the either sides of lens star and the small image with the negative parity is fainter and closer to the lens position. For a large enough impact parameter the position of the negative parity image can be so closer to the lens star that is eclipsed by the lens star. In this case we will miss this image and the overall flux receiving from the source and lens star gets dimmer. Observation of this effect enable us to measure 
the Einstein angle of lens if with photometric or spectroscopic measurements, we can identify the spectral type of the source star and the lens star. We have shown that the a red giant source star and red giant lens star is the favourite system for the observation of this effect and in this case for
less than one year, we would expect to observe the negative-parity image eclipse by the lens star that takes in the order of one day.  During this 
eclipse, the overall flux of lens and source star changes in the order of $10^{-3}$ where with high resolution photometry, we can identify this 
small variation in the flux of microlensed star.

We have shown that a crucial data for prediction of eclipsing time is to identify either the lens star is a red giant or the blending results from the 
background stars in the field. The observations of the high resolution images with lucky camera or observation of the microlensing events 
from the space will help us to resolve the image and find out the origin of blending. The reddish events with the blending factor in the order of $\sim 0.5$ are the suitable systems for the observation of eclipse where in the direction of Galactic bulge it is more likely that both source 
star and lens star to be a red giant.

In order to study the feasibility of observation of this effect, we used a set of microlensing events where all the essential parameters of these 
events for eclipsing time calculation have already been measured. These are mainly high magnification events where the planetary signals in these events is very 
high. Our study show that for the case of lens and source star being red-giant, we can observe this effect and. By combining results from 
the negative-parity eclipsing image with the parallax effect, we can break degeneracy between the lens parameters and measure the 
mass of lens star. The observations of this effect with ongoing microlensing surveys needs identification of the stellar type of source and lens stars that 
can implement in the survey or follow-up telescopes.



\section*{Acknowledgments}
I would like to thank valuable comments of referee. Also, I thank Radek Poleski for his 
useful comments.


\label{lastpage}

\end{document}